\documentclass[conference]{IEEEtran}
\usepackage{cite}
\usepackage{multirow}
\usepackage[table,xcdraw]{xcolor}
\usepackage{amsmath,amssymb,amsfonts}
\usepackage{algorithmic}
\usepackage{graphicx}
\usepackage{textcomp}
\usepackage{url}
\usepackage{bm}
\usepackage{xcolor}
\usepackage{hhline}
\usepackage{bbm}
\usepackage{soul}
\def\BibTeX{{\rm B\kern-.05em{\sc i\kern-.025em b}\kern-.08em
    T\kern-.1667em\lower.7ex\hbox{E}\kern-.125emX}}

\begin{document}
	
\title{Synthetic Time-Series Load Data via Conditional Generative Adversarial Networks}
\author{\IEEEauthorblockN{Andrea Pinceti, Lalitha Sankar, and Oliver Kosut}
\IEEEauthorblockA{School of Electrical, Computer and Energy Engineering, Arizona State University, Tempe, AZ 85287 \\}}
\maketitle
	
\begin{abstract}
A framework for the generation of synthetic time-series transmission-level load data is presented. Conditional generative adversarial networks are used to learn the patterns of a real dataset of hourly-sampled week-long load profiles and generate unique synthetic profiles on demand, based on the season and type of load required. Extensive testing of the generative model is performed to verify that the synthetic data fully captures the characteristics of real loads and that it can be used for downstream power system and/or machine learning applications.
\end{abstract}

	
\section{Introduction}
In recent years, the field of machine learning (ML) has matured to the point where it can provide real value to power system operations; for this reason, a large portion of research work focuses on applying ML to power system applications. Within this new paradigm, the availability of large amounts of real data is crucial. Unfortunately, while power system models of all kinds are readily available, data is a much more scarce resource and the research community must rely on the very few and limited datasets that are publicly available. 

The goal of our project is to develop a mechanism for the generation of synthetic time-series transmission-level load data. Leveraging a proprietary dataset of high resolution measurements from hundreds of phasor measurement units (PMUs) across many years of operation, we can model the behavior of real system loads and subsequently generate realistic-looking data on demand. The focus on bus-level load data is motivated by the fact that loads are one of the main external drivers of power system behaviors; loads depend on phenomena outside of the power system itself (consumer behaviors, weather, etc.). Thus, realistic load profiles can be used as an input to existing power system programs to accurately determine electric quantities such as voltages and currents via dynamic simulation.

In this work, we use a ML technique called conditional generative adversarial network (cGAN) \cite{mirza2014conditional} which represents a powerful and flexible framework for the training of a generative model. Simple GANs \cite{goodfellow2014generative} and conditional GANs have been used in the literature for the generation of PMU voltage data \cite{xiangtian}, renewable energy profiles \cite{baosen}, and residential energy consumption \cite{WANG2020110299} but not for transmission-level load data. We train a cGAN to generate realistic, synthetic week-long time-series load profiles at a resolution of one sample per hour. The generation of synthetic data can be conditioned on labels indicating the season of the year or the type of load profile desired in order to meet the user's specific requirements. 

In the literature, the two main approaches to the generation of synthetic transmission-level load data are to: 1) use prototypical customer load curves and knowledge of the specific geography and demographics to combine them into aggregate loads \cite{Li2020}, or 2) use net or zonal demand and disaggregate it at the bus level based on fixed factors \cite{Grigg}. The main weakness of the first method is the fact that it can only be applied to grid models for which detailed information on the geography of the system and the population served by each load is available: this is required to determine the composition of each load. More importantly, since both methods rely on a limited number of prototypical curves (either zonal or customer level), the resulting synthetic data is limited in diversity and complexity. We follow a newer approach that involves training ML algorithms and generative models on real power system data; the resulting synthetic data can then be mapped to any power system model. Based on this concept, we have proposed in the past a method based on singular value decomposition \cite{pinceti}. The limitations of that work lie in the fact that the synthetic data length was limited to the length of the real dataset used and the generated data only captured main patterns. Using a non-linear model such as conditional generative adversarial networks, we are able to capture more diverse and nuanced load behaviors and overcome the limitations of the previous methods.

\section{Generative Adversarial Networks}\label{GAN}
\subsection{Basic GAN}
Generative adversarial networks are a novel ML framework in which a generative model (or generator) is trained by making it compete against a discriminator. The goal of the generator $G$ is to capture the distribution of the real data $p_r$, while the discriminator $D$ is trained to distinguish the real data from the synthetic data produced by the generator. The generator is trained to learn a mapping $G(\bm{z};\theta_g)$ from a known noise distribution $p_z$ to $p_g$, where $G$ is a differentiable function represented by a multilayer neural network with parameters $\theta_g$ and $\bm{z}$ is a noise vector sampled from $p_z$. Given a data sample $\bm{x}$, the discriminator determines the probability $D(\bm{x}, \theta_d)$ that the sample came from the real data distribution $p_r$ rather than from the generator $p_g$. The training of $D$ and $G$ is represented by a two-player minimax game with the following objective function:
\begin{equation}
    \min_G\max_D  \mathbb{E}_{\bm{x}\sim p_r(\bm{x})}[\log D(\bm{x})]+\mathbb{E}_{\bm{z}\sim p_z(\bm{z})}[\log (1-\ D(G(\bm{z})))]
\end{equation}
Here, the discriminator is maximizing the likelihood of data $x$ when sampled from $p_r$ and minimizing it when sampled from $p_g$, while the generator has the opposite goal of maximizing the likelihood of the samples from $p_g$. The optimal solution is obtained when the discriminator assigns a probability of 0.5 to all samples, meaning that it cannot distinguish between real and generated data.

\subsection{Conditional GAN}
\begin{figure}[h]
    \centering
	\includegraphics[width=9cm,height=4.5cm]{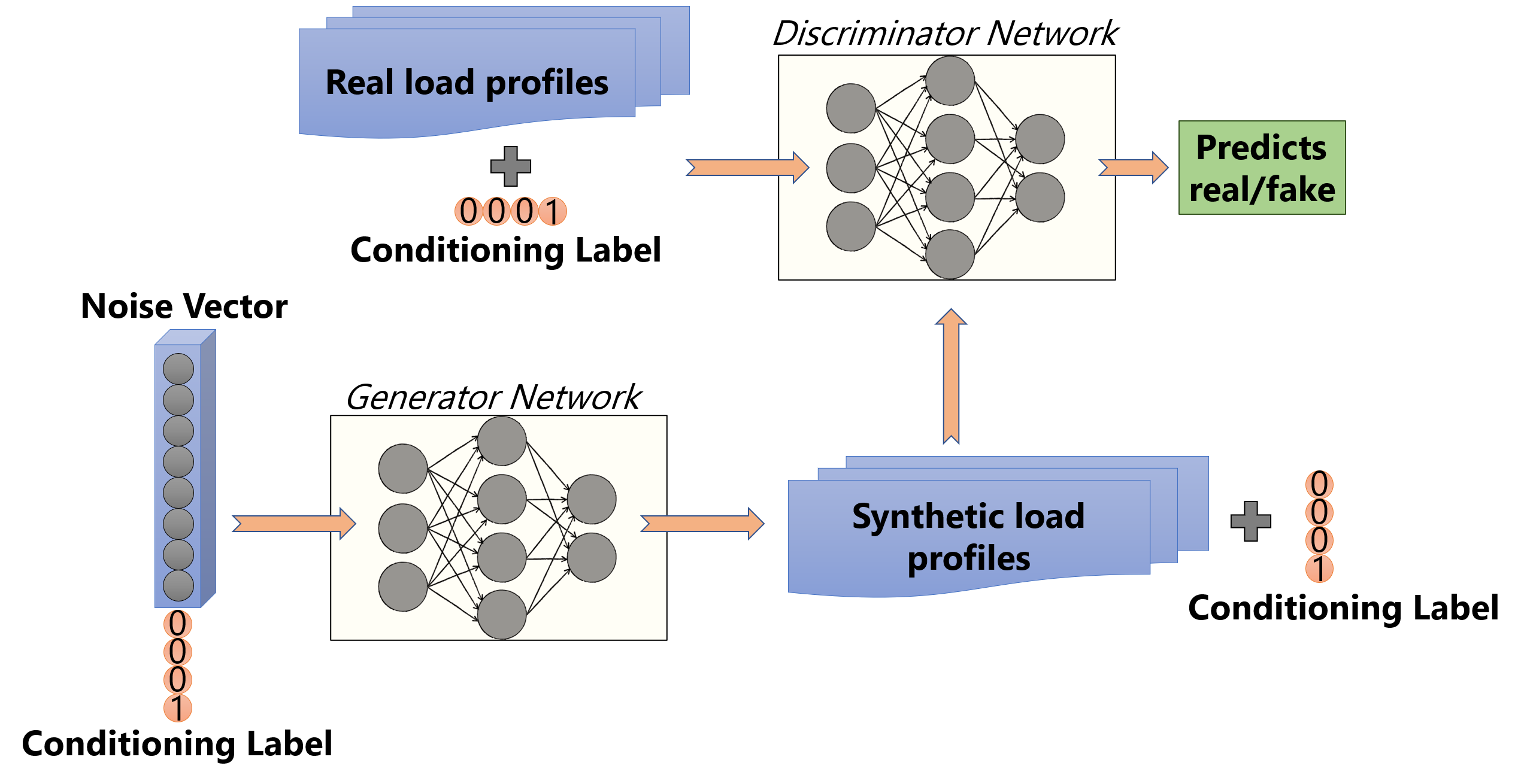}
	\caption{Structure of a conditional GAN.}
	\label{condGAN} 
\end{figure}

Conditional generative adversarial networks (cGAN) are an improvement on the basic GAN framework which allow for a more targeted generation of synthetic data. The conditioning is performed by labelling the real data and then providing this label $\bm{y}$ as a further input to both the generator and the discriminator. By doing this, the generator learns the conditional distribution $p_g$ over $\bm{x}|\bm{y}$ and the generation process can be guided by requesting synthetic data belonging to a specific class. The final structure of a generic cGAN can be seen in Fig.~\ref{condGAN}. 

\section{Dataset Description}
The foundation of this project is a large dataset of real PMU data obtained from a utility in the USA. The data spans two years (2017 and 2018) and about 500 PMUs installed at the transmission level. Based on the system topology and the location of the measurement devices, we identified 12 load buses whose lines are entirely monitored by PMUs. The net injection at these buses represents the load demand and this allowed us to compute the active and reactive power of the 12 loads with a resolution of 30 samples per second. As discussed in the introduction, the focus of this work is the generation of week-long profiles at a resolution of 1 sample per hour for a total of 168 hours. The raw, PMU-speed load data is then downsampled by computing the hourly load average and broken into weeks. When combining all weekly profiles from all 12 loads, the final dataset is a 1158x168 matrix. 
Each week-long profile is normalized by dividing it by the average load over the week; the entire dataset is further normalized between 0 and 1 for the training of the cGAN.

\section{Load Characteristics}
Different factors influence the way system loads behave over time. To appropriately generate realistic synthetic load profiles for a given application, these elements need to be captured and modeled by the GAN. When looking at the week-long time-series data described in the previous section, two main driving factors can be identified: time of the year, and type of load. 

\begin{figure}
    \centering
	\includegraphics[width=9cm,height=5cm]{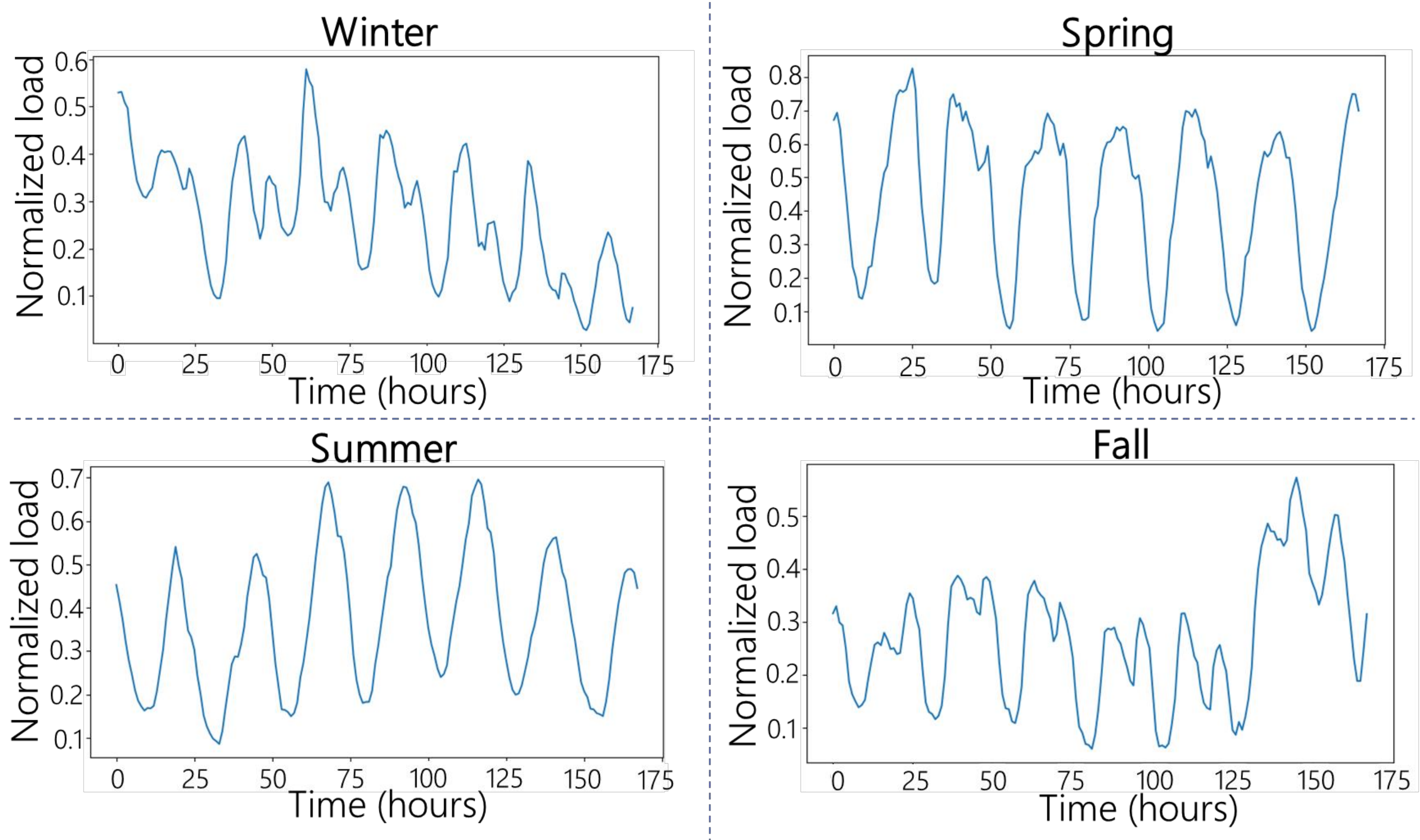}
	\caption{Examples of real profiles; different seasons present different patterns.}
	\label{seasons} 
\end{figure}

The differences between load profiles due to seasonal changes in energy consumption can be easily visualized. Figure~\ref{seasons} shows four week-long profiles for a load, across the four different seasons. In winter and fall, the load pattern presents two daily peaks, one in the morning and one in the afternoon. During spring and especially summer, the load is more regular, with a large spike during the day and dips at night. This type of behavior can be observed in a more or less pronounced manner across all the loads in our dataset. For this reason, the season of the year to which a profile belongs is an important indicator (label) of the expected profile. 

\begin{figure}
    \centering
	\includegraphics[width=9cm,height=4.7cm]{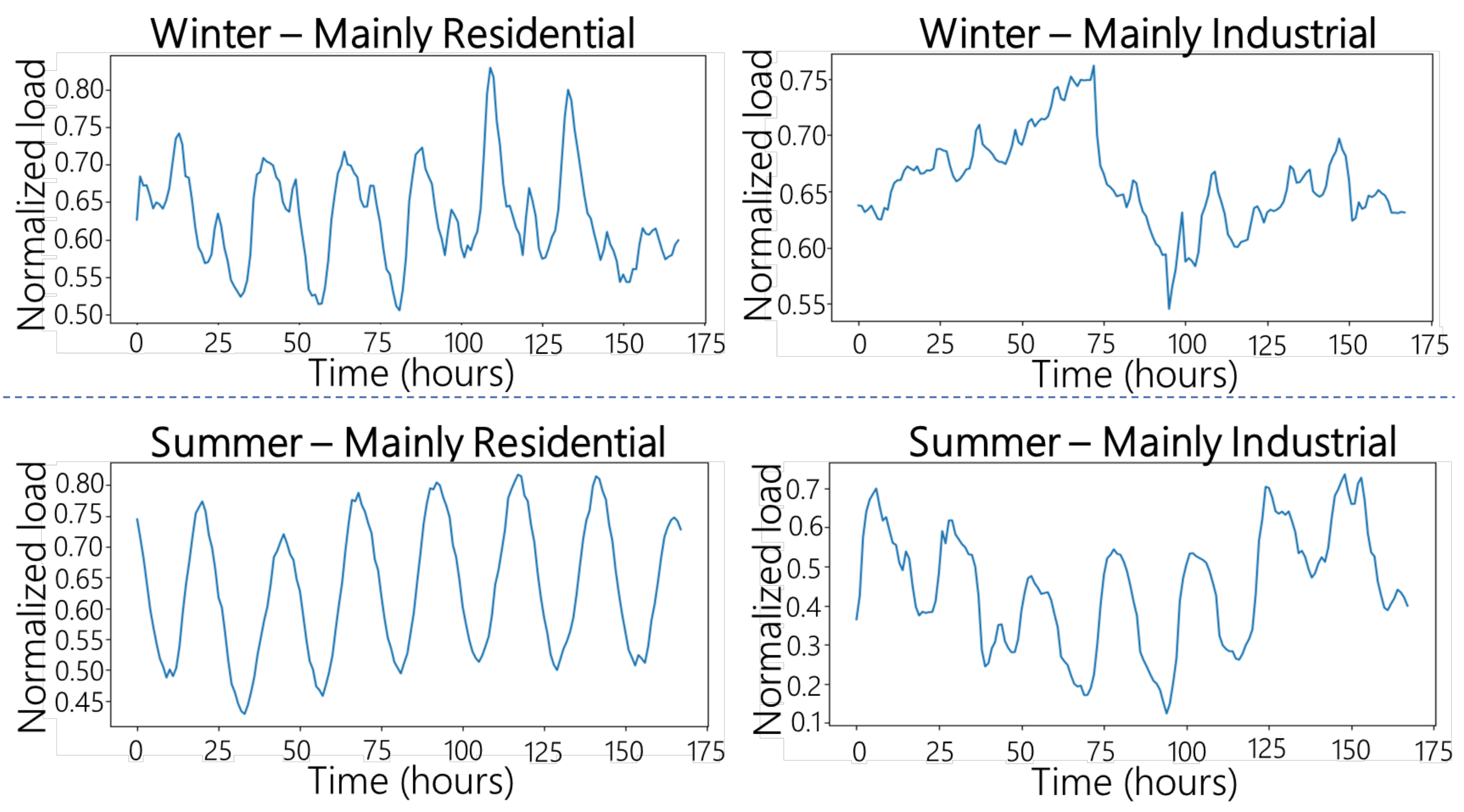}
	\caption{Examples of real load profiles. Top row: mainly residential winter load (left) and mainly industrial winter load (right). Bottom row: mainly residential summer load (left) and mainly industrial summer load (right).}
	\label{types} 
\end{figure}

At the transmission level, each load represents the aggregate demand of one or more distribution feeders. Thus, the behavior of a load is given by the sum of all the customers at the distribution level that it serves. Because of this, the second main factor that determines the temporal profile of a load is its composition in terms of residential, commercial, and industrial portions since each of these types of loads tend to have very distinctive patterns. In our dataset, we have observed two classes of loads with very distinct behaviors: loads that are mainly residential and/or commercial and loads that are mainly industrial. Figure~\ref{types} shows some selected examples: on the top and bottom left are mainly residential loads from winter and summer respectively, while on the right are winter and summer profiles of a mainly industrial load. As we can see, loads that are mainly residential have very regular and predictable patterns, whereas industrial loads do not necessarily follow recognizable daily patterns. We used a $k$-means clustering algorithm to label each load as mainly residential or mainly industrial. When using 3 clusters, two main groups of loads are identified, each containing five and six loads, while one single load is grouped separately. We observed that the loads where the top two factors in terms of percentage composition are residential and commercial are clustered as one, while the loads in which the industrial component is first or second are grouped as another cluster. Thus, for training purposes, six loads are labeled as mainly residential and six as mainly industrial (for more details, we refer the reader to the Appendix found in \cite{Pincetigithub}).

\section{cGAN for Synthetic Load Profiles}

\subsection{cGAN model}
In this section, we will describe the implementation of the cGAN and its training process. Convolutional neural networks (CNNs) are chosen for the discriminator and the generator for their ability to learn multiple spectral properties of the data. While similar in size and complexity, the two models present some differences. 

\begin{figure}
    \centering
	\includegraphics[width=8cm,height=6cm]{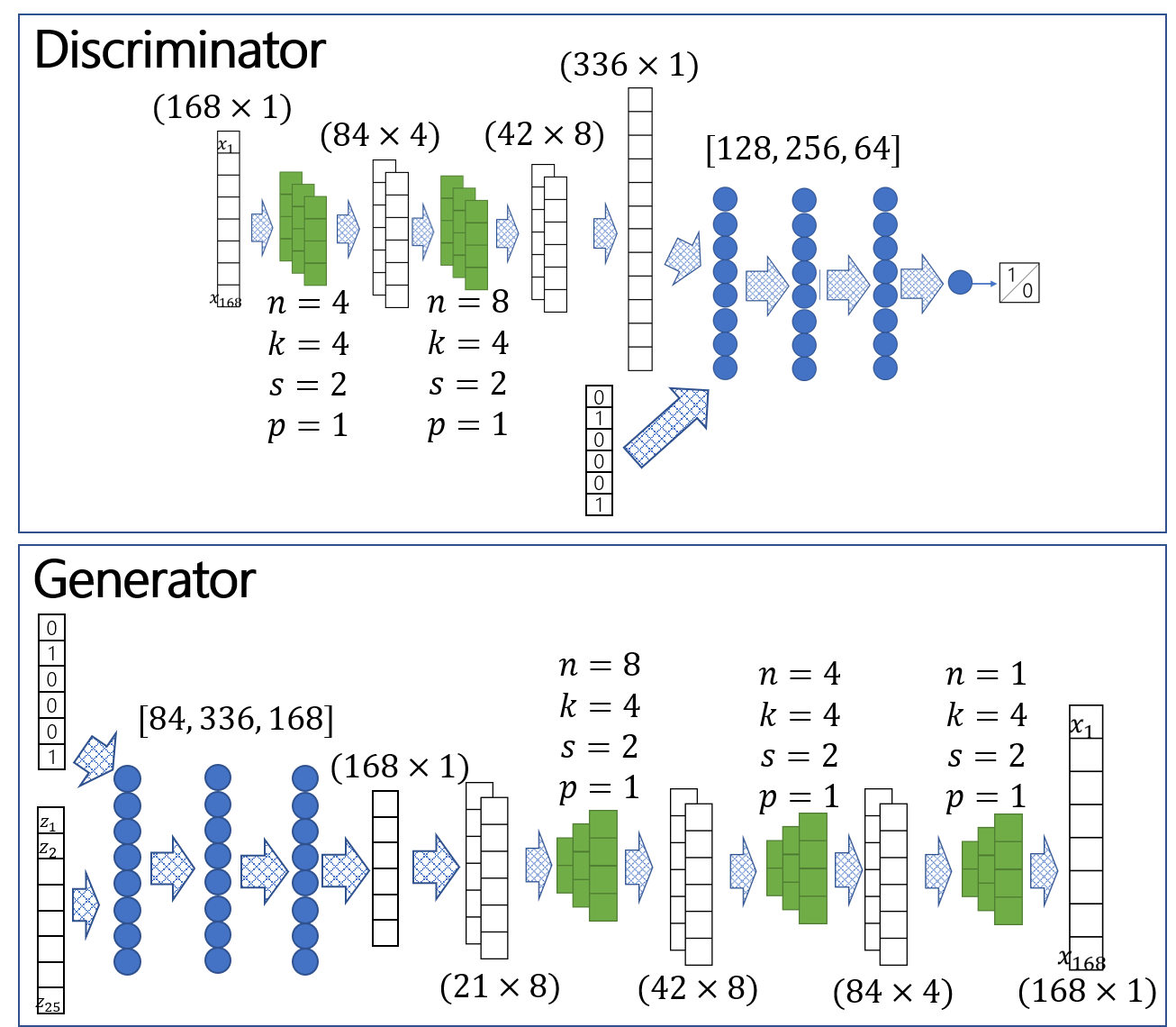}
	\caption{Structure of the cGAN used for the generation of week-long profiles.}
	\label{weekGAN} 
\end{figure}

The discriminator receives two inputs: first, the raw time-series is processed by two convolutional layers, then the flattened output is concatenated to its label and fed to three fully-connected layers. The output of the discriminator is a scalar indicating if a profile is real (1) or fake (0). In the generator, the two inputs are the load label and a 25-dimensional Gaussian noise vector\footnote{The dimension of the noise vector depends on the GAN architectures chosen and the desired output vector length.}. These are concatenated and fed to three fully connected layers and the output is up-sampled via three transposed convolution layers. The final output is a synthetic load profile whose characteristics match the input label. The overall architecture of the cGAN is illustrated in Fig.~\ref{weekGAN}. 

As we have seen in the previous section, two characteristics of loads are used as labels for the conditional GAN: the season and whether a load is mainly residential or industrial. The factors are represented as one-hot encoded vectors, i.e., the seasons are represented via the following four labels: (1 0 0 0) for winter, (0 1 0 0) for spring, (0 0 1 0) for summer, and (0 0 0 1) for fall. Similarly, the load type is encoded as: (0 1) for mainly residential and (1 0) for mainly industrial. 

\begin{figure}
    \centering
	\includegraphics[width=9cm,height=4.3cm]{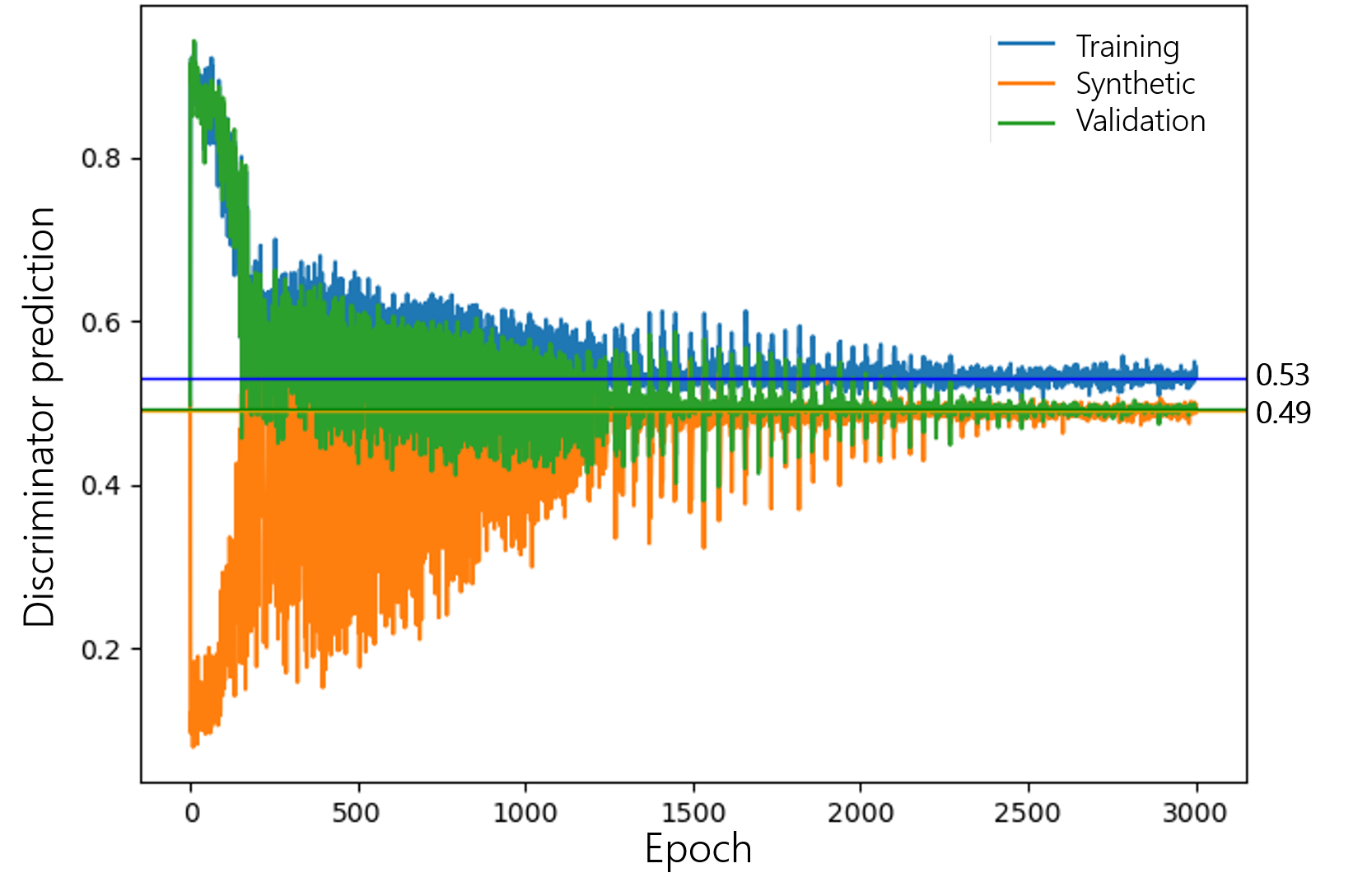}
	\caption{Training progress of the cGAN based on the predictions of the discriminator at each epoch. The blue curve shows the average prediction over a batch of real training profiles. The green curve represents real validation data and the orange one predictions on generated data.}
	\label{trainconv} 
\end{figure}

The training process is performed by iteratively training the discriminator to distinguish between real and generated data and the generator to create realistic-looking profiles. The discriminator is trained twice at every iteration in order to give it an advantage against the generator; this forces the generator to produce better samples. Figure~\ref{trainconv} shows the training progresses as the epochs proceed. The three curves represent the average discriminator prediction at each epoch for three sample datasets: real data used during training (blue), real data never used during training (validation data, green), and fake data created by the generator (orange). We can see that at the beginning the discriminator easily distinguishes between real and fake data, assigning high values to both real datasets and low value to the generated data. As training progresses, the generator improves and the discriminator is unable to differentiate between the two data sources. At around 3000 epochs, the training converges: the discriminator assigns very similar values to all three datasets. It is interesting to notice that some overfitting is happening (the blue curve reaches 0.53) but it is not very significant. More importantly, the discriminator assigns the same values to both the generated data and the validation data; this means the output of the generator matches the real data. The training took approximately 3 hours on a computer with Nvidia RTX2080. Using the trained model to generate synthetic data is extremely fast: 1000 profiles can be generated in less than a second.

\subsection{Data Generation}\label{datageneration}
\begin{figure}
    \centering
	\includegraphics[width=9cm,height=5.5cm]{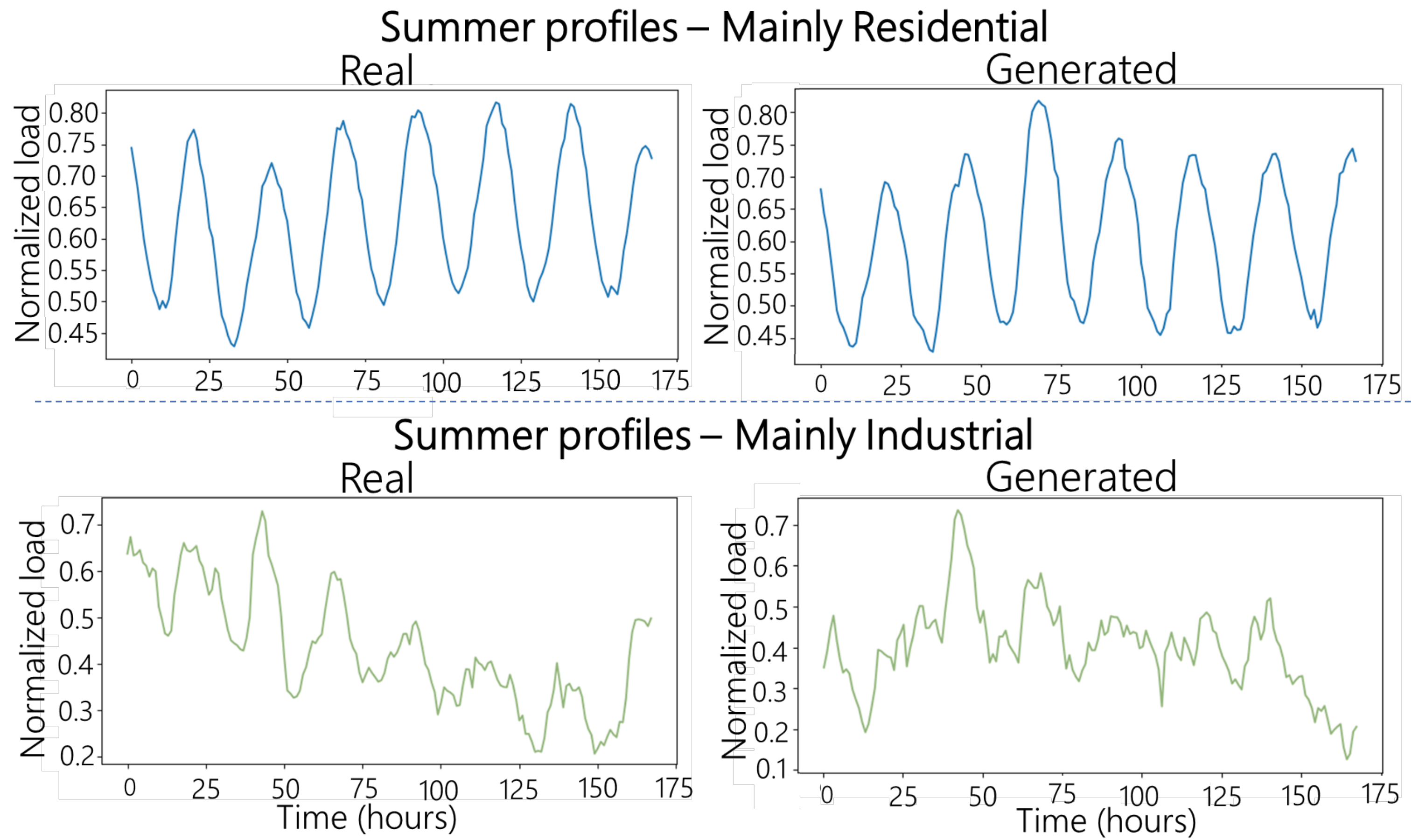}
	\caption{Comparison between some real summer profiles (left) and generated profiles (right). Top: mainly residential load. Bottom: mainly industrial load.}
	\label{001001} 
\end{figure}

Once the training process is terminated, the generator can be used to create any number of synthetic profiles. Based on the required data type, the user only needs to define the appropriate label and generate a noise vector according to the predetermined distribution; feeding these to the generator will result in a synthetic load profile. As an example, in Fig.~\ref{001001} two synthetic summer profiles (right) are compared to two randomly selected real profiles (left) of the same label. The blue profiles (top) correspond to a mainly residential load and the green plots (bottom) to a mainly industrial load. It is important to notice that while the synthetic profiles present all the same characteristics as real data, they do not simply repeat real profiles.

\section{Evaluation of Synthetic Data}
While visual inspection does not yield clear differences between real and synthetic profiles, a quantitative analysis is required to verify that the generator captures all of the characteristics and behaviors present in the real data. In this section, we present multiple tests to validate the quality of the synthetic data. 

\subsection{Wasserstein Distance}
As explained in Section~\ref{GAN}, the goal of the generator is to learn a mapping function from the known noise distribution to the distribution of real data. Training is successful when the distribution of generated data matches that of the real data. Wasserstein distance is a measure of distance between two distributions and it can be used to quantitatively assess how close the distributions of generated and real data are. 

\begin{figure}
    \centering
	\includegraphics[width=9cm,height=3.5cm]{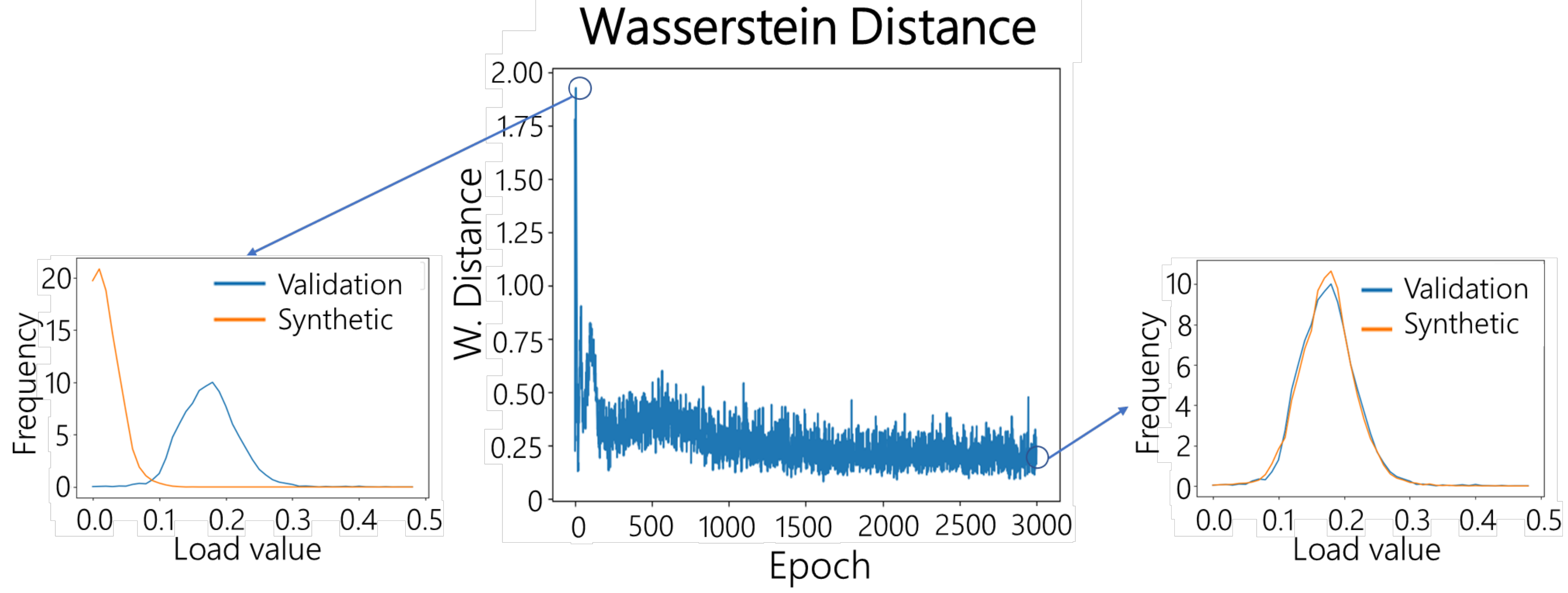}
	\caption{Center plot: Wasserstein distance between real and generated data as a function of epochs. Side plots: comparison between the distribution of real (blue) and generated data (orange) at epoch 0 (left) and epoch 3000 (right).}
	\label{wd} 
\end{figure}

The center plot in Fig.~\ref{wd} shows the Wasserstein distance computed during training at each epoch between a batch of generated data and a batch of validation data. It can be seen that as the training progresses, the distance between the distributions tends to zero. This can be further seen by looking at the two smaller plots on either side. The plots to the left and right show the histograms (empirical distributions) of the real data (blue) and that of the generated data (orange), at epoch 0 (left plot) and at epoch 3000 (right plot), respectively. While initially the two distributions are very different, at the end of training the generated data almost perfectly matches the distribution of real data. 

\subsection{Power Spectral Density}
\begin{figure}
    \centering
	\includegraphics[width=9cm,height=5cm]{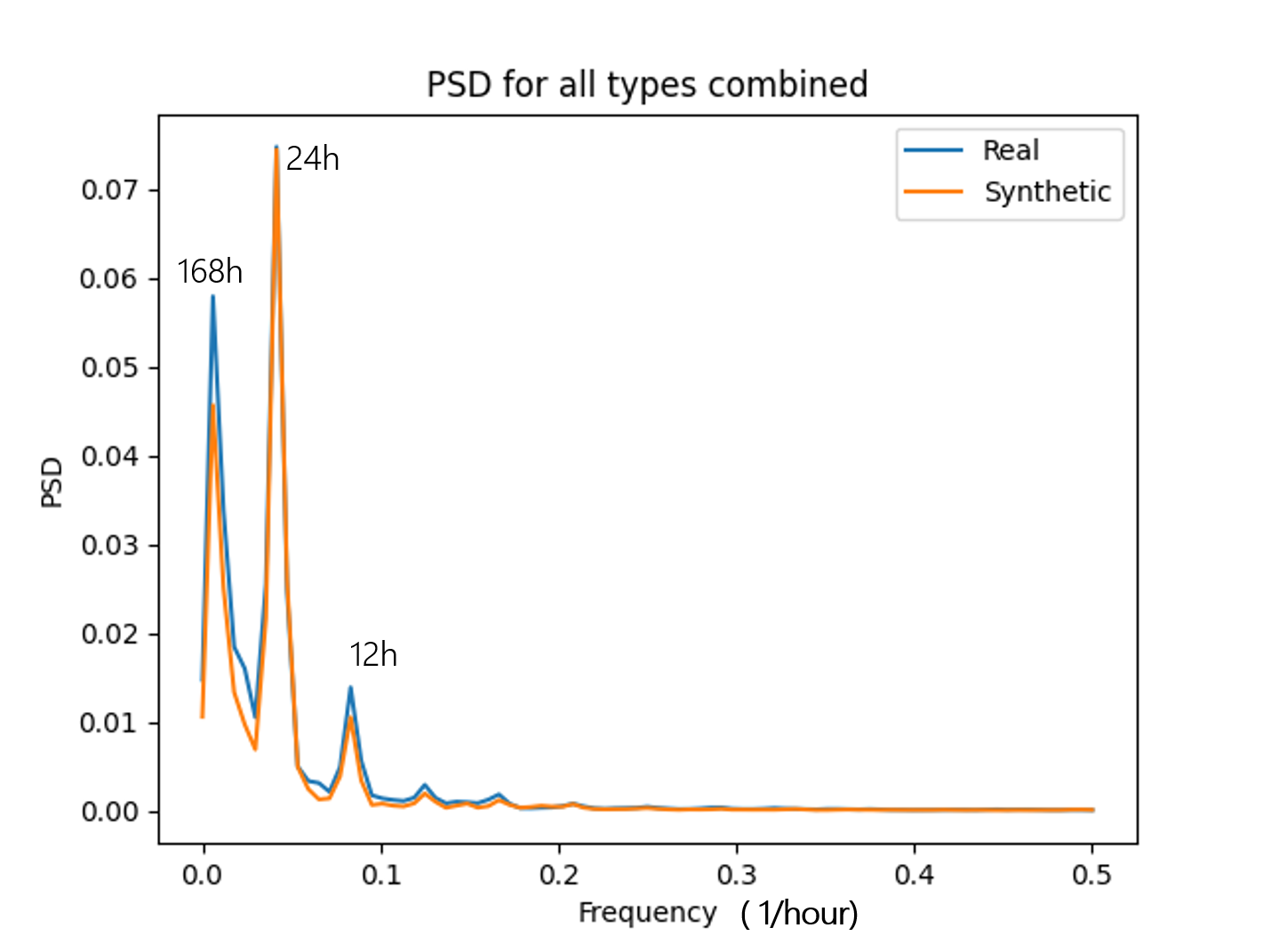}
	\caption{Comparison between the power spectral density of real data (blue) and generated data (orange).}
	\label{psd} 
\end{figure}

An important characteristic of time-series load data is its periodicity. Because loads are tied to the routine and behavior of people, they present different recurring patterns. One way to identify these periodicities is by looking at the power spectral density of the time-series data. Figure~\ref{psd} shows the power spectral density for real data (blue) and generated data (orange). It can be seen that the two plots match very closely, confirming that the generated data captures the periodic behavior of real data. It is also interesting to look at the various peaks that appear in spectral density: in particular, the highest peak, which occurs  at a frequency of 0.04/hour, corresponds to a time period of 24 hours, thus representing the daily load cycle. 

\subsection{Forecasting Application}
The main goal of this paper is to create a mechanism for the generation of realistic synthetic load data that can be used by researchers when real data is either not available or not rich enough. In the next two sections we present the results of two example applications that show that the synthetic data successfully captures the behavior of real data and that it can be used for downstream applications. 

One of the most common uses of time-series load data is the development of the forecasting algorithms needed for power system operations and markets. While many different techniques are used, often in combination, one of the latest advancements in ML-based forecasting is a class of recurrent neural networks called long short-term memory (LSTM). Because feedback loops are present, the LSTM architecture is able to process sequences of data (such as time-series data) maintaining a memory of the previous inputs. To verify the quality of our load data generator, we trained an LSTM on a batch of synthetic data and then tested the learnt model on the real data.  

An LSTM network with three layers and 48 units per layer is trained to predict the value of a load at one point in time, given the previous 48 hours (48 points). This model is trained on two separate datasets independently: synthetic mainly residential summer profiles and synthetic mainly residential fall profiles. Each of the datasets consists of 1200 week-long profiles generated using the trained cGAN according to their respective labels. To evaluate the performance of the LSTM, for each of the two load types, the trained models are used to predict the load values of two batches of profiles: new generated data and real data of the same class. The percentage error between the forecasted value and the actual value is computed for each profile in a batch and the first and second moments are computed. Table~\ref{forecastingtable} summarizes the results of the forecasting test for the two types of loads (summer and fall residential). In both cases we can see that even though the model was trained only on synthetic data, the error when applied to real data is very comparable. In general, this suggests that a user could train a ML model on our synthetic data and be confident that it would still capture the characteristics of real data. 

\begin{table}
\centering
\caption{Comparison of the forecasting error between generated and real load data, for summer and fall residential profiles.}
\begin{tabular}{|l|c|c|c|}
\hline
\multirow{2}{*}{\textbf{Load label}} & \multicolumn{1}{l|}{\multirow{2}{*}{\textbf{Testing data}}} & \multicolumn{2}{c|}{\textbf{Percentage Error}} \\ \cline{3-4} 
 & \multicolumn{1}{l|}{} & \multicolumn{1}{l|}{Mean} & \multicolumn{1}{l|}{Std. Dev.} \\ \hline
\multirow{2}{*}{Summer - Residential} & Synthetic & 4.37 & 5.26 \\ \cline{2-4} 
 & Real & 5.30 & 5.17 \\ \hline
\multirow{2}{*}{Fall - Residential} & Synthetic & 5.82 & 7.10 \\ \cline{2-4} 
 & Real & 5.92 & 5.17 \\ \hline
\end{tabular}
\label{forecastingtable} 
\end{table}

\subsection{Optimal Power Flow}
The synthetic data is also tested to verify that the generated profiles can be correctly mapped to a power system model. One way to check this is to ensure that all the resulting load cases form a feasible AC power flow.

This test is performed by first generating individual, week-long profiles for each load in the Polish test case: this system model has 2383 buses and 1822 loads. Two datasets are generated: one corresponding to a winter week and one for a summer week. Each of these profiles is mapped to the Polish system loads: since the base case of the Polish system is a peak case, the profiles are matched so that the peak of each profile corresponds to the base case value. AC optimal power flow (OPF) is then run on each case corresponding to each of the 168 hours of the week. The results showed that OPF converged in every case to a solution with bus voltages and generator outputs within their predefined limits.

\section{Conclusion}
We have presented a method to generate synthetic transmission load data at a bus level leveraging conditional generative adversarial networks. A user can specify the time of the year and type of load for which to generate time-series load profiles. Extensive testing is performed and we have verified the validity of our method and quality of the generated data. Our trained generative model will be available to researchers to be used for any type of power system and ML application. Moreover, the proposed conditional learning framework can be leveraged for the generation of other datasets highlighting different characteristics, such as the level of penetration of renewables or electric vehicle charging. 
 
Finally, we are working on expanding the generative model to create a tool for the generation of synthetic datasets at any time resolution (from 30 samples/second to a few samples/week) and for any length of time (from a few minutes to multiple years). The trained generative model and an Appendix with extra figures and results can be found at \cite{Pincetigithub}. 

\section{Acknowledgments}
The authors acknowledge Research Computing and Gil Speyer at ASU for providing HPC and storage resources.
This material is based upon work supported by the National Science Foundation under Grant Nos. CNS-1449080 and OAC-1934766 and the Power System Engineering Research Center (PSERC) under projects S-72 and S-87.

\bibliography{main}
\bibliographystyle{ieeetran}

\end{document}